Subject: Source for letter A/51090/LET

\documentstyle[12pt]{article}

\begin{document}

\title{Tunneling in asymmetric double well: instanton calculus}

\author{Alejandro RIVERO \thanks{Dep. F\'{\i}sica 
Teorica, Fac de Ciencias Univ de Zaragoza,
 50009 Zaragoza, Spain} \thanks{email: rivero@cc.unizar.es}}

\maketitle

\begin{abstract}


The level splitting formula of an asymmetric double well potential is
calculated taking into account the multi-instanton contributions
(dilute gas approximation).  Results can be related with known
semiclassical ones obtained with a truncated hamiltonian, and 
the symmetric case is
 easily recovered provided we consider the right limit.

\end{abstract}
ICP: 03.65, 02.30
 

\vskip 1cm

Instanton calculations when applied in Quantum Mechanics result  
powerful methods to build semiclassical approximations, 
see e.g. the lecture from
Coleman \cite{coleman}. Regretly, summation of all the
multinstanton contributions in a problem can be hard work,
and there are only a few examples in the literature, mainly for
highly symmetrical potentials \cite{simons}. When the symmetry
disappears, it is usually need a fully explicit path integral
calculation, with no shorter ways. This could explain
the minimal implementation this technique appears to have
 in more applied physics.

But focusing some less symmetrical potentials, simplified 
summation is also possible, if we assume the series to
 be enough kindly comported.
We show here one possibility for the simplest asymmetric example, a
double well with different values $\omega_0, \omega_1, $  for $V''$
 on the minima 
(see figure 1; think by example on the sextic potential $\lambda
 x^2 (x-2)^4$). Here the fluctuations around the minima have
different contributions, ${1 \over 2} \omega_1> 
{1 \over 2}  \omega_0$, so the n-instanton contribution can not be
directly factorized as usually it is made, taking the integral on 
zero modes apart.

Our method will rearrange the n-instanton integrals in such form
that a recursive definition can be given for all the family, then
obtaining the summation from recurrence equations. Only
 the path between both minimum is calculated, as this is 
enough to show the general strategy. A graphical 
picture is suggested to follow more easily the discussion.

We are going first to calculate the contribution $M_i$ of the multinstanton 
composed of $i+1$ instantons
$\omega_0 \to \omega_1$ and $i$ antinstantons $\omega_1 \to \omega_0$.
Its integral is:
\begin{eqnarray}
\label{start}
M_i&=&N \ K^{2i+1} A^{2i+1} \int_{-T\over 2}^{T\over 2} dt_1
 \int_{t_1}^{T\over 2} dt_2 ... \\ 
\nonumber
&& ...  \int_{t_{2i}}^{T\over 2} dt_{2i+1}
 e^{-{1 \over 2} \omega_0 (t_1-(-{T\over 2}))}
e^{-{1 \over 2} \omega_1 (t_2-t_1)}
e^{-{1 \over 2} \omega_0 (t_3-t_2)}
...
e^{-{1 \over 2} \omega_1 ({T\over 2}-t_{2i+1})}
\end{eqnarray}
where as usual N is the normalization constant, K is the 
one-instanton contribution, and A is the
exponential of the classical action for one instanton.

Note that N and  K are usually calculated (or pulled away) by relating them
to the harmonic oscillator solution. This is unimportant for our discussion
 and we are going to leave out N in the following.
Note also that we have taken equal contributions for instanton and 
antinstanton, as we have put the relevant differences in the integral term.

The integrand in (\ref{start}) being factorizable, we could try to solve the
integral going to complex variable, as it is indicated in any 
handbook (by example, \cite{manual}) but no garantees are given. 
Now if we put $B \equiv KA,\ t \equiv (t_1-t_2)+(t_3-t_4)+...+t_{2i+1}$ we 
can see the
integral in a more suitable form
\begin{equation}
M_i=B^{2i+1} \int_{-T/2}^{T/2} dt_1
 \int_{t_1}^{T/2} dt_2 ...  \int_{t_{2i}}^{T/2} dt_{2i+1}
  e^{\omega_0(T/2+t)} e^{\omega_1(t-T/2)} 
\end{equation}
in which we can rearrange limits and sum up some 
integrals (see again \cite{manual}, 3.3.4), so it rest:
\begin{equation}
\label{instanton}
M_i= B^{2i+1}  e^{-{1 \over 2}  {\omega_0+\omega_1 \over 2} T}
\int_{-T\over 2}^{T\over 2} dt \ e^{{\omega_0-\omega_1 \over 2} t}
{(T/2+t)^i \over i!} {(T/2-t)^i \over i!}
\end{equation}


For the symmetrical well ${\omega_0-\omega_1 \over 2}
\to 0$ and (\ref{instanton}) is simply  Euler' Beta function, which 
evaluates to ${T^{2i+1}
\over 2i+1!}$, the expected zero-mode contribution.  Of course, 
 (\ref{instanton}) could be postulated on physical asumptions,
$(T/2+t)$ and $(T/2-t)$ being the time the instanton stays
 at each vacuum; but
we found useful to point out the derivation process.

Now, we choose not to evaluate this integral,
and we make a simultaneous study of integrals of the kind
\begin{equation}
I(n,m)= B^{n+m+1}  e^{-{1 \over 2}  {\omega_0+\omega_1 \over 2} T}
\int_{-T/2}^{T/2} dt \ e^{\delta t}
{(T/2+t)^n \over n!} {(T/2-t)^m \over m!}
\end{equation}

Integrating by parts we can give the following recursive
definition for all the family:
\begin{eqnarray}
I(0,0)=&&{B\over \delta} [e^{{1\over 2} \delta T}-e^{-{1\over 2} \delta T}] \\
I(n,0)=&&{B\over \delta}[e^{{1\over 2} \delta T} {(BT)^n \over n!}-I(n-1,0)] \\
I(0,m)=&&{B\over \delta} [I(0,m-1)
-e^{-{1\over 2} \delta T} {(BT)^m \over m!}] \\
I(n,m)=&&{B\over \delta} [I(n,m-1)-I(n-1,m)]
\end{eqnarray}
which can be seen in a pictorial form by putting the integrals
in a triangle (figure 2), such that each integral
is obtained by substracting the two above it and putting an
additional factor $B/\delta$.

So, $I(n,m)$ can be directly calculated by inspecting the 
triangle, counting and weighting the number of paths from
each term in the sides. The result is:
\begin{eqnarray}
\label{int.resuelta}
I(n,m)&=&e^{-{1\over 2} \omega_0 T} \sum_{i=0}^{i=n} {m+n-i\choose m}
(-1)^{n-i} ({B\over \delta})^{n+m-i+1} {(BT)^i \over i!} + \\
\nonumber
&+& e^{-{1\over 2} \omega_1 T} \sum_{j=0}^{j=m} {m+n-j\choose n}
(-1)^{n+1} ({B\over \delta})^{n+m-j+1} {(BT)^j \over j!}
\end{eqnarray}


We could again be tempted, now of using (\ref{int.resuelta}) to
sum all the multinstanton contributions. But it results
that we can avoid it by newly inspecting the triangle and fixing
our attention on sums of columns, which must, as everything
in these combinatorial triangles, own some interesting properties.
 
Let $S(n,m)$ be the sum of a column $\sum_{i=0} I(n+i,m+i)$, and let
$S_j^{\pm}(n,m)$ be the coefficient of $S(n,m)$ associated to the term
$  { (BT)^j \over j!} e^{\pm\delta T/2}$. Grafically (see
 figure 3) we find some relations between sums:
\begin{eqnarray}
S_i(n,m) =&& {B \over \delta} [S_i(n,m-1) -S_i(n-1,m)] \\
S_i^+(n,m)=&&S_{i+1}^+(n+1,m) \\
n<i \Rightarrow S_i^+(n,m)=&&S_i^+(i,m+(i-n)) \\
\label{main.rule}
S_i^+(i,m)=&&{B \over \delta} [S^+_i(i,m-1)-S_i^(i,m+1)]= \\
\nonumber
   =&&{B \over \delta} [S^+_{i-1}(i-1,m-1)-S_{i+1}^(i+1,m+1)] 
\end{eqnarray}

In particular, this last rule (\ref{main.rule}), when applied to the 
central column, lets us to formulate a recurrence equation for the
coefficients $a_i^\pm \equiv S^\pm_i(i,i)$ of the series of powers 
in $ { (BT)^j \over j!}$.
Symply put, we get:
\begin{equation}
a_{i+1}^\pm=a_{i-1}^\pm \mp {\delta\over B} a_i^\pm 
\end{equation}
which corresponds to the series produced by a lineal combination of two 
exponentials $C_+ e^{\alpha_+} +C_- e^{\alpha_-}$.

Let us to make it explicitly for the $S^+$ terms. In this case
the coefficients in the exponentials are:
\begin{equation}
\alpha_\pm=-{\delta \over 2 B} \pm \sqrt{({\delta \over 2 B})^2 +1}
\end{equation}
and it rests to fix $C_+,C_-$; we can made it
from the two first terms of the serie, which obviusly fulfill:
\begin{eqnarray}
a_0=&&C_+ +C_- \\
a_1=&&-(C_++C_-) {\delta \over 2 B} 
+(C_+-C_-)  \sqrt{({\delta \over 2 B})^2 +1}
\end{eqnarray}
So, we finally need to sum two series,
\begin{eqnarray}
a_0=\sum_{i=0} {2i \choose i} (-1)^i \left({B \over \delta}\right)^{2i+1}
   ={B \over \delta} \sum_{i=0} {1\over 2}..{2i-1 \over 2} (-1)^i 
			{\left(2B/\delta\right)^{2i} \over i!} \\
a_1=\sum_{i=0} {2i-1 \choose i-1} (-1)^{i-1} \left({B \over \delta}\right)^{2i}
   =\sum_{i=1} {1\over 2}..{2i-1 \over 2} (-1)^{i-1} 2^{2i-1} 
			{(2B/\delta)^{2i} \over i!}
\end{eqnarray}
but we can acomplish it by simply browsing across any
handbook and remembering the expansion of $(1+x)^{-1/2}$,
and we obtain:
\begin{eqnarray}
a_0=&&{1 \over 2 \sqrt{1+({\delta \over 2B})^2}} \\
a_1=&&{1 \over 2} - {1 \over 2 \sqrt{1+({2B \over \delta})^2}}
\end{eqnarray}

And from here we get $C_-=0$ and $C_+=
{1 \over 2 \sqrt{1+({\delta \over 2B})^2}}$.

Now, for the $S^-$ part we see, by symmetry of the triangle,
that the calculus is similar, and we get the same coefficients
$C_+,C-, $ but interchanged and with
a sign changed.

The final result is, thus,
\begin{equation}
\sum_{i=0} M_i={1 \over 2 \sqrt{1+({\delta \over 2B})^2}}
[e^{-E_+T}-e^{-E_-T}]
\end{equation}
with 
\begin{equation}
\label{result}
E_\pm={\omega_0 + \omega_1 \over 4} \mp \sqrt{{\delta^2 \over
4} + B^2}
\end{equation}
And expanding, we have that the gap between the first two
levels is:
\begin{equation}
\Delta=\sqrt{{(\omega_1-\omega_0)^2 \over 4} + 4 K^2 e^{-2 S_{inst}}}
\end{equation}

As a consistency check,  we observe that when we go back
to the symmetrical case, $\delta \to 0$,
equation (\ref{result}) gives us the well known formula for the
energy splitting. And of course, when the instanton
contribution is negligible, $B\to 0$, we get two separated wells
with same energy levels. 

We can compare this result with the "ancient" semiclassical method,
merely diagonalization of the truncated hamiltonian matrix 
\begin{equation}
H=\left( \begin{array}{cc} 
\omega_0 / 2 & B'	\\
B'& \omega_1 / 2 \end{array} \right)
\end{equation}
and we obtain again (\ref{result}), only replacing $B$ by $B'$.
Then also in the asymmetrical case we can use the instanton
method to estimate the barrier penetration factor $B'$.

The method can be complicated in two directions: we can add
more minima with different curvature, which force us to leave
off the plain-triangle and obscures the method; or we can use
it to treat potentials with more minima but only two curvatures,
as by example the polynomical $x^6$ triple well, which add some 
instantons more to calculate, but doesn't add other different
summations.  

Finally, note that this operation
method implies some rearrangements of power series, some care of the
convergency conditions must be put when working on a more general case.
Apart from this, a strong asymmetry can force the breakdown described
by Jona Lasinio et al., see \cite{jona},\cite{annals}.


We would like to thank J. Esteve and J. Casahorran by useful discussions
and pointers. This job has been supported in part with funds provided
by CICYT (Spain). The
author must acknowledge grant AP9029093359.



\begin{thebibliography}{20}
\bibitem{coleman} S. COLEMAN, ``Aspects of Symmetry'', Cambridge Univ. Press, 
1985 
\bibitem{manual} A.P PRUDNIKOV et al.,  ``Integrals and Series'', vol. I,
Gordon and Breach Science Publishers.
\bibitem{simons} B. SIMON,
Semiclassical analysis of low lying eigenvalues II: Tunneling,
 {\em Annals of Mathematics}, {\bf 120} (1984) 84 
\bibitem{annals} F. CESI, 
Non-symmetric Double Well and Euclidean Functional Integral,
 {\em Annals of Physics}, {\bf 205} (1991) 318
\bibitem{jona} G. JONA-LASINIO et al, 
New approach to the semiclassical limit of quantum mechanics, 
{\em Communications in Mathematical Physics}, {\bf 80} (1981) 223. 

\end{thebibliography}
\end{document}